\documentclass[aps,prl,twocolumn,showpacs,preprintnumbers,amsmath,amssymb,10pt]{revtex4-1} 
\usepackage{amsmath,graphicx,subfig,xcolor,verbatim}
\usepackage{diagbox}
\usepackage{tikz}
\usepackage{braket}
\usepackage{float}
\usetikzlibrary{calc,decorations.pathmorphing}

\usepackage[dvipsnames]{xcolor}
\usepackage{hyperref}

\hypersetup{
    colorlinks=true,
    linkcolor=CadetBlue,     
    citecolor=CadetBlue,       
    urlcolor=Periwinkle      
}

\newcommand{\be}{\begin{equation}}
\newcommand{\ee}{\end{equation}}
\newcommand{\bea}{\begin{eqnarray}}
\newcommand{\eea}{\end{eqnarray}}
\newcommand\restr[2]{{\left.\kern-\nulldelimiterspace#1\vphantom{\big|}\right|_{#2}}}

\newcommand{\beq}{\begin{equation}} 
\newcommand{\eeq}{\end{equation}}

\begin{document}
\title{Ising surface defects can get dirty}
\author{Ant\'onio Antunes$^\phi$, Apratim Kaviraj$^\Phi$, Baishali Roy$^\Phi$} 
\affiliation{
 $^\phi$Laboratoire de Physique de l'\'Ecole Normale Sup\'erieure, Universit\'e  PSL, CNRS, Sorbonne Universit\'e, Universit\'e  Paris Cit\'e, 24 rue Lhomond, F-75005 Paris, France\\
 $^\Phi$Department of Physics, Indian Institute of Technology - Kanpur, Kanpur 208016, India}
\begin{abstract}
Real critical systems, such as uniaxial ferromagnets in the 3d Ising universality class, are constrained by boundaries and subject to random couplings. We consider the Wilson-Fisher fixed point in $4-\epsilon$ dimensions subject to a random magnetic field localized on a two-dimensional surface, which becomes co-dimension 1 in the physical $\epsilon\to1$ limit. Using the replica method for the disordered field, we find that the ordinary boundary condition is stable under disorder but also discover a non-trivial ``dirty" boundary condition which can be reached by tuning the disorder strength or the local temperature. We also investigate the logarithmic structure of the defect spectrum and how it emerges via the replica formalism.
\end{abstract}
\maketitle
\nopagebreak
\, \\
{\bf Introduction:}
Defects and disordered couplings are two powerful non-local probes of critical systems described by conformal field theories (CFTs). On the one hand, $p-$dimensional conformal defects \cite{Billo:2016cpy,Lemos:2017vnx,Lauria:2017wav,Lauria:2018klo}, preserving the maximal residual symmetry
\begin{align}
    G_{\rm{Eucl}} &= SO(p+1,1)\times SO(d-p)\,,  \nonumber\\
    G_{\rm{Lor}}  &= SO(p,2)\times SO(d-p)\,,
\end{align}
can describe, for example spatially localized interactions in euclidean statistical systems, or timelike impurities in Lorentzian/quantum systems \cite{Cuomo:2021rkm,Cuomo:2021kfm,Cuomo:2022xgw}. These non-local probes have their own associated conformal Hilbert space, describing defect local operators $\hat{\mathcal{O}_i}$, their scaling dimensions $\hat{\Delta}_i$ and OPE coefficients $\hat{C}_{ijk}$, as well as defect operator expansion (DOE) coefficients $\mu_i^A$, which allow to rewrite bulk operator insertions $\mathcal{O}^A$ as infinite sums of defect operators $\hat{\mathcal{O}_i}$. All of these numbers constitute an independent set of CFT data which is not uniquely fixed from the knowledge of the bulk theory: Indeed, a given bulk CFT has potentially infinitely many consistent conformal defects.

On the other hand, disordered couplings, which modify the theory by the addition of a spatially varying coupling $h(x)$ (drawn from a random distribution) to the formal action
\begin{equation}
    S_h=S_{\rm{CFT}} + \int d^dx\,h(x) \,\mathcal{O}(x)\,,
\end{equation}
lead to an RG flow which can modify the IR universality class, following Harris' famous criterion\,\cite{Harris}. In general, such spatially varying couplings lead to non-unitary CFTs, which often have logarithmic structures. In fact, for a given disorder realization $h(x)$, the couplings break translational invariance and sometimes global symmetries, which typically are restored after averaging over disorder realizations. Quite remarkably, there can even be new symmetries that emerge in the IR of the RG flow. 
An interesting case of emergent symmetries is that of Parisi-Sourlas supersymmetry (PS SUSY) which arises at an IR fixed point of `random field' models, e.g.  Ising model with a random magnetic field coupled to Ising spins. This SUSY fixed point is in fact a nonunitary CFT exhibiting logarithmic properties \cite{Parisi:1979ka,Kaviraj:2019tbg,Tarjus:2019qoq,Kaviraj:2020xjj,Kaviraj:2021qii,Kaviraj:2022xsp,Trevisani:2024mgi,Tarjus:2024mop}. Non-unitarity is an intrinsic property of disorder averaged theories that makes the study of such CFTs rather challenging \cite{Wiese:LectureNotesStatFieldTheory,Komargodski:2016auf,Rychkov:2023rgq}.


However, the notion of conformal defects discussed above suggests a way of probing the theory with disorder while maintaining some amount of unitarity: Restricting the disorder to a $p-$dimensional submanifold! This will keep our unitary bulk universality class untouched, while introducing a generically non-unitary (and possibly logarithmic) conformal defect. Aside from its use as a theoretical device to probe non-unitary physics, localized disordered couplings are quite natural experimentally: they describe ``dirty'' surfaces or edges of a material. Such systems were first considered in \cite{Ghosh:2025lzb} in the context of Parisi-Sourlas supersymmetry, and their kinematics was studied in great detail in \cite{Shimamori:2025afk} which also established many explicit results in the case of a free scalar CFT in the bulk. A related setup was explored in \cite{yang2026boundarycriticalitynishimorimulticritical}  with tensor-network methods.

In this work, we will consider the classic example of the Wilson-Fisher fixed point in $4-\epsilon$ dimensions, subject to a random magnetic field localized on a two dimensional defect, with the hope of studying the 3d Ising universality class when $\epsilon\to 1$. This is therefore just a defect-localized version of the aforementioned random field Ising model (RFIM) \cite{JLCardy_1991}. In our case, the defect becomes an interface in the 3d theory, but it is expected that it actually describes a boundary condition, as such perturbative defects are often known to factorize onto the two sides \cite{Popov:2025cha}. Working in the replica formalism, we find a new disordered fixed point, which, however turns out to be RG unstable with respect to the ordered ordinary defect fixed point, but can be reached by tuning the disorder strength and a local ordered coupling. Then, we compute anomalous dimensions of linear and bilinear defect operators, emphasizing the organization of the spectrum into logarithmic conformal multiplets which emerge in the replica limit. \\ 
\\
{\bf Disorder setup:}
Consider a massless interacting scalar field theory in the presence of a disorder surface defect, given by the action: 
\begin{equation}\label{disordersetup}
S \!= \!\!\int \!d^d x \left[ \frac{1}{2} (\partial \Phi)^2 \!+\! \frac{\lambda}{4!} \Phi^4 \right] 
   \!+ \!\!\int \!d^2 \tau\, \!\!\left[h(\tau) \phi\!+\!g\, \phi^2\right]\,,
\end{equation}
where the surface defect coupling $h(\tau)$ is a disorder coupling randomly chosen from a Gaussian distribution $\mathcal{P}(h) \propto e^{-\frac{1}{2H} \int d^2 \tau \, h^2(\tau)}$ with a zero mean,  $\overline{h(\tau)} = 0$ and a variance $\overline{h(\tau) h(\tau')} = H \delta(\tau - \tau')$ ensuring zero spatial  correlation.

Throughout this paper we will use
$\Phi$ and $\phi$ to denote the bulk fields and the ones restricted to the 2d surface respectively, and decompose the coordinates as $$(\{x_1, \ldots, x_{d-2}\}, \{\tau_1, \tau_2\}) \equiv (\vec{x},\vec{\tau}), $$ where $\vec{x}\in\mathbb{R}^{d-2}$ label the direction transverse to the defect, and $\vec{\tau}\in \mathbb{R}^2$ are defect coordinates.

Our observables, such as correlation functions, will be computed in the framework of quenched disorder averaging \footnote{Usually, for a disorder interaction, two kinds of averaging can be done: Annealed and Quenched  depending on the kind of impurities present. We will be interested in the latter, which is usually done when the defects are not in thermal equilibrium. For this the expectation is at first computed for a fixed  disorder coupling $h$ and then the averaging over the disordered coupling is considered. 
(see eg. \cite{Cardy1996} for a detailed discussion)}.
E.g. for a collection of defect fields  $A(\phi)$ one can consider the averaged correlator
\begin{equation}\label{avg}
\overline{\langle A(\phi) \rangle} = 
\int \mathcal{D}h\, \mathcal{P}(h) 
\frac{\int \mathcal{D}\phi\, A(\phi)\, e^{-S[\phi, h]}}{Z_h}
\end{equation}
where 
\begin{equation}
Z_h = \int \mathcal{D}\phi\, e^{-S[\phi, h]}\,, 
\end{equation}
However, a direct evaluation of this is not straightforward as the partition function $Z_h$ fluctuates with disorder. The expectation therefore involves averaging over a ratio of disorder dependent quantities which do not factorize into simpler averages, rendering the disorder averaging non-trivial. \\ \\
{\bf The replica trick:}
As reviewed in \cite{Kaviraj:2019tbg}, one can use the method of replicas \cite{Cardy1996,Cardy:2013rqg} to rewrite the averaged correlator \eqref{avg} as follows:
\begin{align}\label{replicaEq2}
\overline{\langle A(\phi) \rangle} 
&= \lim_{N \to 0} 
\int \mathcal{D}h\, P(h)
\int \mathcal{D}\vec{\phi}\, 
A(\phi_1) 
e^{-\sum_{i=1}^N S[\phi_i, h]} \nonumber \\ 
&=\lim_{N \to 0} 
\int \mathcal{D}\vec{\phi}\, 
A(\phi_1)\, e^{-S_N[\vec{\phi}]} \,. 
\end{align}

where
\begin{align}\label{replicaAction}
S_N[\Phi] &= 
\int d^d x 
\left[
    \sum_{i=1}^{N} \left(  \frac{1}{2} (\partial \Phi_i)^2 + \frac{\lambda}{4!} \Phi_i^4  \right)
   \right] \nonumber \\&- \int d^2\tau \Big[\frac{1}{2} H 
    \big( \sum_{i=1}^{N} \phi_i \big)^2+g\sum_{i=1}^N\phi_i^2\Big]
\end{align}
and $\vec{\phi}=\{\phi_i\}$ with the index $i$ ranging from 1 to $N$. Here, $\phi_1$ represents the physical observable fields and $\phi_2,\dots, \phi_N$ are the auxiliary replica fields, introduced to handle $Z_h^{-1}$ in the path integral.

The main idea of the trick is to trade $Z_h^{-1}$ for $Z_h^{N-1}/Z_h^{N}$ in \eqref{avg}. Writing the factor $Z_h^{N-1}$ as $Z_h^{N-1}=\prod_{i=2}^n\int \mathcal{D}\phi_i e^{-S[\phi_i,h]}$ and assuming that the analytic continuation to $N\rightarrow0$ limit is valid the rhs of \eqref{avg} can be rewritten without a denominator since the factor $Z_h^N\rightarrow 1$ in the $N\rightarrow0$ limit.
 This allows us to perform the integral over the disorder coupling before finally taking $N\rightarrow0$ limit in \eqref{replicaEq2}.
Note that, we also have a coupling $g$, which is a defect mass term, or a localized change in temperature. Tuning this additional coupling will be important to find non-trivial fixed points. 

In disordered theories, interesting observables also include averages of products of correlation functions. The average of  $\langle A(\phi)\rangle\langle B(\phi)\rangle\cdots$ can be obtained in the replica method from the correlator $\langle A(\phi_1)B(\phi_2)\cdots\rangle$ at $N\to 0$. E.g. the quantity $\overline{\langle\phi(\tau)\rangle\langle\phi(0)\rangle}$ is obtained from $\langle\phi_1(\tau)\phi_2(0)\rangle$ computed in the action \eqref{replicaAction}. We will see later on the importance of taking different combinations of replicas in correlation functions. 
\\ \\
{\bf Finding the ``dirty'' fixed points:}
We will find it convenient to redefine the couplings and work with the replica action: 
\begin{align}
S_N &= \int d^d x 
\left[
    \sum_{i=1}^{N} \frac{1}{2} (\partial_\mu \Phi_i)^2 
    + \frac{ \lambda}{4!} \sum_{i=1}^{N} \Phi_i^4
\right] \nonumber\\ 
&- \int d^2 \tau 
\left[
    \frac{h_1}{2}  \sum_{i=1}^{N} \phi_i^2
    + \frac{h_2}{2}  \sum_{{i\neq j}}^{N} \phi_i \phi_j
\right]
\end{align}
where, $h_1$ and $h_2$ are two defect couplings which couple fields from the same or different replica respectively. We are interested in studying the RG flow triggered by these couplings, and the IR fixed points the defect flows to, while sitting at the interacting Wilson-Fisher fixed point in the bulk.

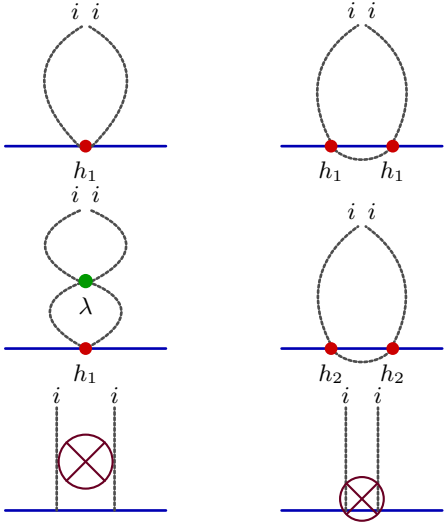
\begin{figure}[H]

\centering
\begin{tikzpicture}[scale=0.85, line cap=round, line join=round]

  \tikzset{
    defect/.style={blue!70!black, line width=1pt},
    prop/.style={black!70, densely dotted, line width=1pt},
    redpt/.style={circle, fill=red!80!black, inner sep=1.7pt},
    greenpt/.style={circle, fill=green!60!black, inner sep=1.9pt},
    ctbubble/.style={purple!55!black, line width=0.8pt}
  }

  \def\xsep{4.3}
  \def\ysep{-3.15}

  \begin{scope}[shift={(0,0)}]
    \draw[defect] (-1.25,0) -- (1.25,0);
    \node[redpt,label=below:$h_1$] at (0,0) {};

    \draw[prop]
      (-0.10,0) .. controls (-0.95,0.85) and (-0.70,1.48) .. (-0.06,1.86);
    \draw[prop]
      ( 0.10,0) .. controls ( 0.95,0.85) and ( 0.70,1.48) .. ( 0.06,1.86);

    \node at (-0.16,2.10) {$i$};
    \node at ( 0.16,2.10) {$i$};
  \end{scope}

  \begin{scope}[shift={(\xsep,0)}]
    \draw[defect] (-1.25,0) -- (1.25,0);

    \coordinate (A) at (-0.48,0);
    \coordinate (B) at ( 0.48,0);

    \draw[prop] (A) .. controls (-0.24,-0.28) and (0.24,-0.28) .. (B);

    \draw[prop]
      ($(A)+(0.03,0)$) .. controls (-0.95,0.90) and (-0.62,1.52) .. (-0.06,1.88);
    \draw[prop]
      ($(B)+(-0.03,0)$) .. controls ( 0.95,0.90) and ( 0.62,1.52) .. ( 0.06,1.88);
\node[redpt,label=below:$h_1$] at (A) {};
    \node[redpt,label=below:$h_1$] at (B) {};

    \node at (-0.16,2.12) {$i$};
    \node at ( 0.16,2.12) {$i$};
  \end{scope}

  \begin{scope}[shift={(0,\ysep)}]
    \draw[defect] (-1.25,0) -- (1.25,0);

    \draw[prop]
      (0,0) .. controls (-0.82,0.42) and (-0.66,0.83) .. (0,1.05)
            .. controls ( 0.66,0.83) and ( 0.82,0.42) .. (0,0);

    \draw[prop]
      (-0.10,1.05) .. controls (-0.92,1.46) and (-0.70,1.90) .. (-0.06,2.16);
    \draw[prop]
      ( 0.10,1.05) .. controls ( 0.92,1.46) and ( 0.70,1.90) .. ( 0.06,2.16);
 \node[greenpt,label=below:$\lambda$] at (0,1.05) {};
\node[redpt,label=below:$h_1$] at (0,0) {};
    \node at (-0.16,2.40) {$i$};
    \node at ( 0.16,2.40) {$i$};
  \end{scope}

  \begin{scope}[shift={(\xsep,\ysep)}]
    \draw[defect] (-1.25,0) -- (1.25,0);

    \coordinate (A) at (-0.48,0);
    \coordinate (B) at ( 0.48,0);

    \draw[prop] (A) .. controls (-0.24,-0.28) and (0.24,-0.28) .. (B);

    \draw[prop]
      ($(A)+(0.03,0)$) .. controls (-0.92,0.84) and (-0.63,1.50) .. (-0.06,1.90);
    \draw[prop]
      ($(B)+(-0.03,0)$) .. controls ( 0.92,0.84) and ( 0.63,1.50) .. ( 0.06,1.90);
 \node[redpt,label=below:$h_2$] at (A) {};
    \node[redpt,label=below:$h_2$] at (B) {};
    \node at (-0.16,2.14) {$i$};
    \node at ( 0.16,2.14) {$i$};

  \end{scope}

  \begin{scope}[shift={(0,1.8*\ysep)}]
    \draw[defect] (-1.25,0) -- (1.25,0);
    \draw[prop] (-0.45,0) -- (-0.45,1.6);
    \draw[prop] ( 0.45,0) -- ( 0.45,1.6);
    \draw[ctbubble] (0,0.76) circle[radius=0.42];
    \draw[ctbubble] (-0.30,1.06) -- (0.30,0.46);
    \draw[ctbubble] (-0.30,0.46) -- (0.30,1.06);
    \node at (-0.45,1.8) {$i$};
    \node at ( 0.45,1.8) {$i$};
  \end{scope}

  \begin{scope}[shift={(\xsep,1.8*\ysep)}]
    \draw[defect] (-1.25,0) -- (1.25,0);
    \draw[prop] (-0.25,0) -- (-0.25,1.6);
    \draw[prop] ( 0.25,0) -- ( 0.25,1.6);
    \draw[ctbubble] (0,0.18) circle[radius=0.34];
    \draw[ctbubble] (-0.24,0.42) -- (0.24,-0.06);
    \draw[ctbubble] (-0.24,-0.06) -- (0.24,0.42);
      \node at (-0.25,1.8) {$i$};
    \node at ( 0.25,1.8) {$i$};
  \end{scope}
\end{tikzpicture}
\caption{Defect Feynman diagrams for same replica. The crosses indicate counter terms.}
\label{fig:samerepdiag}
\end{figure}

\begin{figure}[H]

\centering
\begin{tikzpicture}[scale=0.85, line cap=round, line join=round]

  \tikzset{
    defect/.style={blue!70!black, line width=1pt},
    prop/.style={black!70, densely dotted, line width=1pt},
    redpt/.style={circle, fill=red!80!black, inner sep=1.7pt},
    greenpt/.style={circle, fill=green!60!black, inner sep=1.9pt},
    ctbubble/.style={purple!55!black, line width=0.8pt}
  }

  \def\xsep{4.3}
  \def\ysep{-3.15}

  \begin{scope}[shift={(0,0)}]
    \draw[defect] (-1.25,0) -- (1.25,0);
    \node[redpt,label=below:$h_2$] at (0,0) {};

    \draw[prop]
      (-0.10,0) .. controls (-0.95,0.85) and (-0.70,1.48) .. (-0.06,1.86);
    \draw[prop]
      ( 0.10,0) .. controls ( 0.95,0.85) and ( 0.70,1.48) .. ( 0.06,1.86);

    \node at (-0.16,2.10) {$i$};
    \node at ( 0.16,2.10) {$j$};
  \end{scope}

  \begin{scope}[shift={(\xsep,0)}]
    \draw[defect] (-1.25,0) -- (1.25,0);

    \coordinate (A) at (-0.48,0);
    \coordinate (B) at ( 0.48,0);

    \draw[prop] (A) .. controls (-0.24,-0.28) and (0.24,-0.28) .. (B);

    \draw[prop]
      ($(A)+(0.03,0)$) .. controls (-0.95,0.90) and (-0.62,1.52) .. (-0.06,1.88);
    \draw[prop]
      ($(B)+(-0.03,0)$) .. controls ( 0.95,0.90) and ( 0.62,1.52) .. ( 0.06,1.88);

    \node[redpt,label=below:$h_1$] at (A) {};
    \node[redpt,label=below:$h_2$] at (B) {};
    \node at (-0.16,2.12) {$i$};
    \node at ( 0.16,2.12) {$j$};
  \end{scope}

   \begin{scope}[shift={(0,\ysep)}]
    \draw[defect] (-1.25,0) -- (1.25,0);
    \draw[prop] (-0.25,0) -- (-0.25,1.6);
    \draw[prop] ( 0.25,0) -- ( 0.25,1.6);
    \draw[ctbubble] (0,0.18) circle[radius=0.34];
    \draw[ctbubble] (-0.24,0.42) -- (0.24,-0.06);
    \draw[ctbubble] (-0.24,-0.06) -- (0.24,0.42);
      \node at (-0.25,1.8) {$i$};
    \node at ( 0.25,1.8) {$j$};
  \end{scope}
  \begin{scope}[shift={(\xsep,\ysep)}]
    \draw[defect] (-1.25,0) -- (1.25,0);

    \coordinate (A) at (-0.48,0);
    \coordinate (B) at ( 0.48,0);

    \draw[prop] (A) .. controls (-0.24,-0.28) and (0.24,-0.28) .. (B);

    \draw[prop]
      ($(A)+(0.03,0)$) .. controls (-0.92,0.84) and (-0.63,1.50) .. (-0.06,1.90);
    \draw[prop]
      ($(B)+(-0.03,0)$) .. controls ( 0.92,0.84) and ( 0.63,1.50) .. ( 0.06,1.90);
 \node[redpt,label=below:$h_2$] at (A) {};
    \node[redpt,label=below:$h_2$] at (B) {};

    \node at (-0.16,2.14) {$i$};
    \node at ( 0.16,2.14) {$j$};
  \end{scope}

\end{tikzpicture}
\caption{Defect Feynman diagrams for different replicas.}
\label{fig:difrepdiag}
\end{figure}
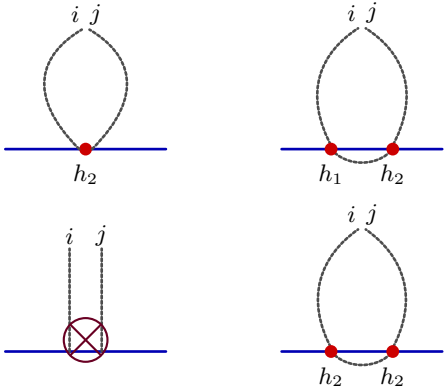

To compute the RG equations, we consider one-point correlation function of the bulk composite operator $\langle [\Phi_i \Phi_j](\vec{x}, \vec{\tau}=0) \rangle$ located at a finite distance from the defect surface \cite{Giombi:2023dqs,Trepanier:2023tvb,Raviv-Moshe:2023yvq}. We will derive the RG equations and beta functions by demanding the finiteness of the renormalized composite operator at $\epsilon\rightarrow0$ 
\begin{equation}
\langle [\Phi_i \Phi_j](\vec{x}, 0) \rangle \equiv \text{finite}
\end{equation}
i.e, by adding necessary counterterms to absorb the loop singularities (poles in $\epsilon$) order by order in our perturbative analysis. This leads us to consider two different situations: \\
\textbf{Case I:} when $i = j$ and we consider contribution to the composite operator $\langle [\Phi_i \Phi_i](\vec{x}, 0) \rangle$. The contributing Feynman diagrams up to one loop, including bulk and defect counterterms,  are given in Fig. \ref{fig:samerepdiag}. 
Notice that there is a contribution from the bulk $\Phi^4_i$ coupling.
\ \\
\textbf{Case II:} when  $i \neq j$, the diagrams contributing to  $\langle [\Phi_i \Phi_j](\vec{x}, 0) \rangle$ are given in Fig. \ref{fig:difrepdiag}.

We immediately note that no bulk coupling appears at one loop in the second case. Using standard dimensional regularization and the minimal subtraction scheme and considering the bulk coupling $\lambda$ at criticality, we find the relation between the bare \footnote{Note that all the coupling  written so far are the bare couplings. The defect coupling $h_1,h_2$ in the replica action have the superscript $(0)$, i.e., $h_1^{(0)}$ and $h_2^{(0)}$, which we have not written explicitly before.} and renormalized couplings to be
\begin{align}\label{bareRenormalizedRelation}
h_1^{(0)} &= M^{\epsilon} \left( h_1 - \frac{1}{\pi \epsilon} \left(  \frac{h_1^2}{2} + \frac{h_2^2}{2} (N-1) -\frac{\lambda h_1 }{16 \pi} \right) \right), \\\label{bareRenormalizedRelation2}
h_2^{(0)} &= M^{\epsilon} \left( h_2 - \frac{1}{\pi \epsilon} \left( h_1 h_2 + (N - 2) \frac{h_2^2}{2} \right) \right).
\end{align}
where $M$ is the renormalization scale. Introducing a renormalization matrix 
$h_i^{(0)} = M^\epsilon h_j Z_{ij}$,
the beta functions can be read off as
\begin{equation}
\beta_{h_i} = -\epsilon \, h_i -  \sum_k h_j\, \beta_{g_k}\,\frac{\partial \ln Z_{ij}}{\partial g_k},
\end{equation}
where $g_i \in \{h_1,h_2,\lambda\}$. At one loop, the entries to $Z$ are read off from \eqref{bareRenormalizedRelation}, and one finds 
\begin{align}
\beta_{h_1} &= -h_1 \epsilon - \frac{h_1^2}{2\pi} - \frac{ h_2^2( N-1)}{2\pi} + \frac{\lambda^* h_1}{16 \pi^2}, \label{beta1} \\
\beta_{h_2} &= -h_2 \epsilon -\frac{h_1 h_2}{\pi}  - \frac{ h_2^2}{2\pi} (N - 2). \label{beta2}
\end{align}
where we evaluated $\lambda$ at its nontrivial fixed point $
\lambda_* = \frac{16\pi^2}{3}\,\epsilon$. 
Solving these equations simultaneously, and taking the replica index $N \to 0$ limit, we find the following fixed points:
\begin{align}
(h_1^\ast, h_2^\ast) &= (0, 0) \\
(h_1^\ast, h_2^\ast) &= \left(-\frac{4 \pi  \epsilon }{3}, 0\right) \\
(h_1^\ast, h_2^\ast) &= \left(-\frac{3 \pi  \epsilon }{2} ,  -\frac{\pi  \epsilon }{2} \right) \end{align}
The first fixed point is the UV trivial defect. The second one is an ordered fixed point (same for all $N$) corresponding to an ordinary-type boundary condition, already discovered in \cite{Giombi:2023dqs,Trepanier:2023tvb,Raviv-Moshe:2023yvq}. The third and final fixed point is new and intrinsically disordered and will be the center of our attention for the remaining of the paper. For generic $N$ there is also a fourth non-trivial fixed point but it goes to infinity in the $N\to0$ limit. It would be interesting to investigate the nature of the associated phase, but it is reasonable to expect that it is associated to a $\mathbb{Z}_2$ breaking boundary condition - an extraordinary phase, in analogy with the standard phase diagram of the Ising model with boundary interactions. 

\begin{figure}
    \centering
\includegraphics[scale=0.34]{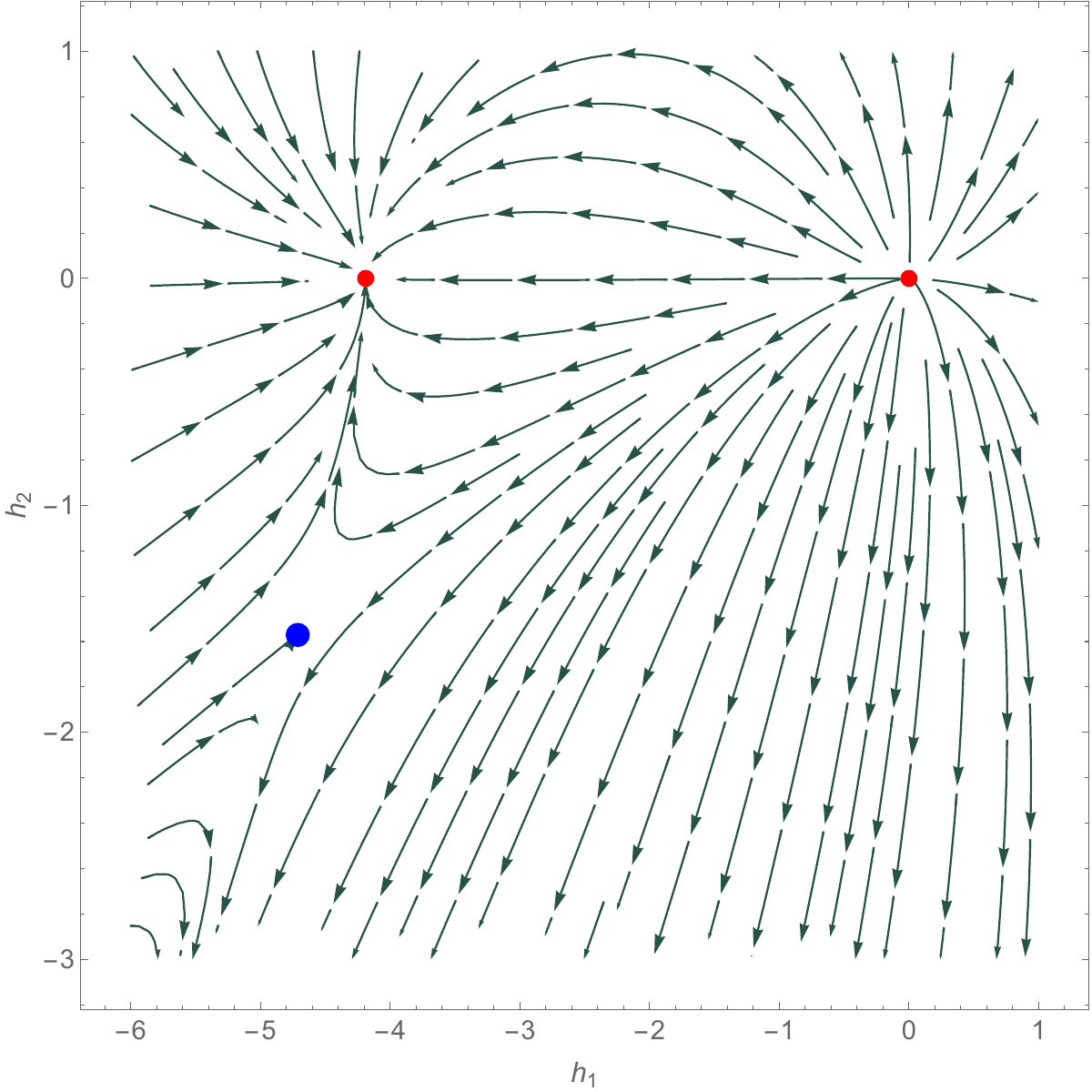}
    \caption{Flow plot for the surface defect RG flow. The two ordered fixed points are denoted by a red dot, and the disordered one by a blue dot.}
\label{fig:flow}
\end{figure}

By computing the stability matrix $\partial \beta_{h_i}/\partial h_j$, we can read off the anomalous dimensions of the two bilinear singlet deformations $\sum_i\phi_i^2$ and $\sum_{i\neq j} \phi_i\phi_j$. This allows us to infer that the stable fixed point is actually the ordered one, as can be seen in the flow plot of Figure \ref{fig:flow}.
For the disordered defect, the corresponding eigenoperators are both a mixture of the two singlets with their scaling dimensions given by 
\begin{align}\label{DeltaAB}
    \Delta_A&=2+\frac{1+\sqrt{7}}{6}\epsilon\,, \quad \Delta_B =2+\frac{1-\sqrt{7}}{6}\epsilon\,.
\end{align}
In particular, this fixed point has mixed stability: while it will not generically control the IR physics, it is possible to reach this non-trivial fixed point by tuning $H/g$ from \eqref{disordersetup}, which can be achieved, for example by tuning the surface temperature with respect to the bulk critical temperature. \\ \\
{\bf Renormalization of defect fields:} 
We begin by computing the anomalous dimensions of the fundamental defect field $\phi^i$. Under the $S_N$ replica symmetry, the operator decomposes into two irreducible representations
\begin{equation}\label{phi-irrep}
    \phi^T=\sum_i \phi_i\,, \quad \tilde{\phi}_i=\phi_i-\frac{1}{N} \phi^T\,,
\end{equation}
which are the singlet/trace and the $N-1$ dimensional fundamental/vector representation. We will first compute the anomalous dimensions for general $N$ and discuss the interesting structure of the $N\to 0$ limit in the next section. 

\vskip 1cm
\begin{figure}[h]
    \centering
\begin{tikzpicture}[scale=0.9]

\begin{scope}[yshift=-1cm]

\draw[blue, thick] (0,0) -- (3.6,0);

\draw[thick] (0.9,0) arc[start angle=180,end angle=0,radius=0.9];

\fill (0.9,0) circle (1.2pt);
\fill (2.7,0) circle (1.2pt);
\node[below] at (0.9,0) {$i$};
\node[below] at (2.7,0) {$i$};
\node at (1.8,-0.55) {$a_0$};


\end{scope}

\begin{scope}[yshift=-3cm]

\draw[blue, thick] (0,0) -- (3.6,0);

\draw[thick] (0.6,0) arc[start angle=180,end angle=0,radius=0.6];

\draw[thick] (1.8,0) arc[start angle=180,end angle=0,radius=0.6];

\fill[red] (1.8,0) circle (2pt);

\fill (0.6,0) circle (1.2pt);
\fill (3.0,0) circle (1.2pt);
\node[below] at (0.6,0) {$i$};
\node[below] at (3.0,0) {$i$};
\node[below] at (1.8,0) {$h_1$};

\node at (1.8,-0.90) {$a_1$};


\end{scope}

\begin{scope}[xshift=4.8cm,yshift=-3cm]

\draw[blue, thick] (0,0) -- (3.6,0);

\draw[thick] (0.6,0) arc[start angle=180,end angle=0,radius=0.6];

\draw[thick] (1.8,0) arc[start angle=180,end angle=0,radius=0.6];

\fill[red] (1.8,0) circle (2pt);

\fill (0.6,0) circle (1.2pt);
\fill (3.0,0) circle (1.2pt);
\node[below] at (0.6,0) {$i$};
\node[below] at (3.0,0) {$j$};
\node[below] at (1.8,0) {$h_2$};

\node at (1.8,-0.90) {$b_0$};


\end{scope}
\end{tikzpicture}
 \caption{The diagrams contributing to $\langle \phi_i(\tau_1)\phi_j(\tau_2)\rangle$.}
    \label{fig:phiphi2point}
\end{figure}
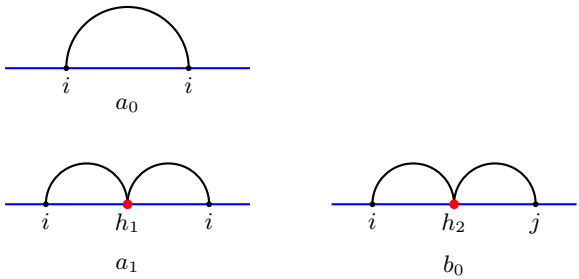

At first, we note that up to first order in the couplings, the two point defect correlation function has the following contributions
\begin{align}\label{ddcorr1}
&\braket{\phi_i(\tau_1)\phi_j(\tau_2)}=\delta_{ij} (a_0+a_1(h_1))+(1-\delta_{ij})b_0(h_2) 
\end{align}
where, $a_0$ is the free propagator term $G(\tau_1;\tau_2)=(4\pi^2|\tau_1-\tau_2|^2)^{-1}$ contributing to the same replica indexed defect correlators, and $a_1$ and $b_0$ are the first order corrections in defect coupling (see Fig. \ref{fig:phiphi2point}). They can be obtained from the integral \cite{Giombi:2023dqs,Trepanier:2023tvb,Raviv-Moshe:2023yvq}: 
\begin{align}
I&=\int d^2 \vec{\tau} G(\vec{\tau}_1;\vec{\tau}) G(\vec{\tau};\vec{\tau}_2)\nonumber\\
&=\int  d^2 \vec{\tau} \int \frac{d^d p_1}{(2\pi)^d} \frac{d^d p_2}{(2\pi)^d} \frac{e^{i \mathbf{q}_1. (\vec{\tau}_1-\vec{\tau})} e^{i \mathbf{q}_2. (\vec{\tau}-\vec{\tau}_2)}}{p_1^2p_2^2}\nonumber\\
&=\mathcal{N}_{1,d-2}^2 \int \frac{d^2 \mathbf{q}_1}{(2\pi)^2}  \frac{e^{i \mathbf{q}_1. (\vec{\tau}_1-\vec{\tau}_2)}}{(\mathbf{q}_1^2)^{4-d}}=\frac{\mathcal{N}_{1,d-2}^2 C_{2,4-d}}{((\vec{\tau}_1-\vec{\tau}_2)^2)^{d-3}}
\end{align}\\ \sloppy
where  $\mathcal{N}_{a,b}=(4\pi)^{-\frac{b}{2}}\Gamma(a)^{-1}\Gamma(a-\frac{b}{2})$ and $C_{r,s}=\Gamma(\frac{r}{2}-s)/(4^s \pi^{r/2} \Gamma(s)) $. For $d=4-\epsilon$ in $\epsilon\rightarrow0$, this becomes $I=(4 \pi^3 \epsilon\  \vec{\tau}_{12}^2)^{-1}$. With correct combinatorics, we find the following:
\begin{align}
&a_0=\frac{1}{4\pi^2|\tau_1-\tau_2|^2}\,, a_1(h)=\frac{h \, a_0}{\pi \epsilon}\,, b_0(h)=\frac{h \, a_0}{2\pi \epsilon}\,.
\end{align} 
We can obtain the two point function of the singlet $\braket{\phi^T(\tau_1)\phi^T(\tau_2)}$ from \eqref{ddcorr1} by summing over $i,j$. The 2-point function of the vectors $\braket{\tilde{\phi}_k(\tau_1)\tilde{\phi}_l(\tau_2)}$ is obtained from \eqref{ddcorr1} by contracting the tensor $\tilde{\delta}_{ki}\tilde{\delta}_{jl}$ where $\tilde{\delta}_{ij}:=\delta_{ij}-\frac{1}{N}$.

Taking the $Z$-factors
corresponding to the renormalization of fields $\phi^T(\tau)$ and $\tilde{\phi}_i(\tau)$ to be $Z_{\phi}=(1+c_1)$ and $Z_{\tilde{\phi}}=(1+c_2)$ and demanding the renormalized two point functions  $\braket{\phi^R(\tau_1)\phi^R(\tau_2)}$ and $\braket{\tilde{\phi}^R_i(\tau_1)\tilde{\phi}^R_j(\tau_2)}$ (where $\phi^R$ is the renormalized $\phi^T$) to be finite, we find
$$c_1=\frac{h_1}{2\pi \epsilon}+\frac{h_2(N-1)}{4\pi \epsilon}\,, \quad c_2=\frac{h_1}{2\pi \epsilon}-\frac{h_2}{4\pi \epsilon}\,. $$
Therefore, the scaling dimensions of $\phi(\tau)$ and $\tilde{\phi}_i(\tau)$ are,
\begin{align}
    &\Delta_\phi=\frac{d-2}{2}-\frac{h_1+\frac{h_2 (N-1)}{2}}{2\pi}|_{h_i=h_i^*}\,,\nonumber\\
    &\Delta_{\tilde{\phi}}=\frac{d-2}{2}+\frac{(\frac{h_2}{2}-h_1)}{2 \pi}|_{h_i=h_i^*}\,.
\end{align}
The above results hold for any $N$ and for any of the defect fixed points. Below we specialise to the disordered one and examine its special features. \\ \\ 
{\bf Logarithmic structure via replica limit:} Let us now take the results of the previous section and carefully study the $N\to0$ limit. It is well known (see e.g.  \cite{Cardy:2013rqg, Hogervorst:2016itc}) that a logarithmic CFT emerges in this limit. The defining feature of such theories is that the dilatation generator $D$ cannot be fully diagonalized and has a Jordan normal form, i.e. there exist sets of operators $\{\mathbb{O}_a\}$, for each of which $[D,\mathbb{O}_a]={\bf \Delta}_{ab}\mathbb{O}_b$ with the Jordan block  
\be
{\bf \Delta}=\left(
\begin{array}{cccc}
 \Delta  & 1 & 0 & \cdots \\
 0 & \Delta  & \ddots & \ddots \\
 \vdots & \ddots & \ddots & 1 \\
0 & \cdots & 0 & \Delta  \\
\end{array}
\right)\,.
\ee
If ${\bf \Delta}$ is a $2\times 2$ Jordan block then we have a rank 2 logarithmic multiplet $(\mathbb{O}_1, \mathbb{O}_2)$ with 2-point function 
\be
\langle\mathbb{O}_a(x)\mathbb{O}_b\rangle \sim x^{-2 \Delta } \left(
\begin{array}{cc}
 \log (x) & 1 \\
 1 & 0 \\
\end{array}
\right)_{ab}\,.
\ee
We refer the reader to \cite{Hogervorst:2016itc} for further details.   \\

In a replica theory, logarithmic modules are expected to emerge \cite{Cardy:2013rqg} from linear combinations of operators whose scaling dimensions become degenerate at $N\to 0$. It is not a priori clear which operators in which representations will recombine into the log multiplets. But a straightforward strategy  is to identify the ones that have identical dimensions verify whether they have a logarithmic 2-point function.

To this end it is useful to classify them into $S_N$ irreducible representations \cite{Gurarie:1993xq,Read:2007ph,Dubail:2010im,Vasseur_2012}. Let us start from the single defect field  operators $\phi_i$. The  singlet and the vector from  \eqref{phi-irrep} have the following 2-point functions
:
\begin{align}\label{mulcorr1}
    &\braket{\phi^T(\tau)\phi^T(0)}=\frac{N c_0}{\tau^{2\Delta_{\phi}}}\,,\\ &\braket{\tilde{\phi}_i(\tau)\tilde{\phi}_j(0)}=\frac{(\delta_{ij}-\frac{1}{N})c_0}{\tau^{2\Delta_{\tilde{\phi}}}}\,,\nonumber\\ &\braket{\phi^T(\tau)\tilde{\phi}_i(0)}=0\,,\nonumber
\end{align}
where $c_0=1/(4\pi^2)$. Now
consider
2-point correlators involving the field $\phi_1$. Since $\phi_i=\tilde{\phi}_i+\frac{1}{N}\phi^T$ \footnote{Strictly speaking one should write $\phi_i=\tilde{\phi}_i+\frac{M^{\Delta_{\tilde\phi}-\Delta_\phi}}{N}\phi^T$, but we ignore the scale dependence for brevity. The scale will drop out in the end anyway.},  the 2-point function of $\phi_1$ with itself becomes  
\begin{equation}\label{corf1}
    \braket{\phi_1(\tau) \phi_1(0)}=(1-\frac{1}{N})\frac{\ c_0}{\tau^{2 \Delta_{\tilde{\phi}}}}+\frac{1}{N}\frac{ \ c_0}{\tau^{2 \Delta_{\phi}}}\,.
\end{equation}
while that of $\phi_1$ and $\phi^T$ is  
\begin{equation}\label{corf2}
\braket{\phi_1(\tau)\phi^T(0)}=\frac{c_0}{\tau^{2 \Delta_{\phi}}}.
\end{equation}
Now $\Delta_{\phi}=\Delta_{\tilde{\phi}}-Nh^*_2/(4\pi)$ as $N\rightarrow0$, the $1/N$ singularities cancel and we obtain a logarithmic 2-point function 
\begin{align}\label{log-phi}
    \lim_{N\rightarrow0}\braket{\phi_1(\tau) \phi_1(0)}=\frac{2 c_0 h_2^* \log \tau}{4 \pi \tau^{2\Delta_{\phi}}}
\end{align}
where $\Delta_\phi=1+\frac{\epsilon}{8}$.
This way we identify $\phi_1,\phi^T$ as the logarithmic partners from \eqref{corf1} and \eqref{corf2}.  \\

Logarithmic multiplets also appear from composite operators. Consider bilinears of the fields $\phi_i$. There are five such operators belonging to three irreps of $S_N$. Writing explicitly, we have the two singlets
\begin{equation}
    S_1=(\phi^T)^2\,,\, S_2=\sum_i (\tilde{\phi}_i)^2\,,
\end{equation}
two vectors (fundamentals/standard representation)
\begin{equation}\label{vectors}
    V_{1,i}= \phi^T \,\tilde{\phi}_i\,,    V_{2,i}=(\tilde{\phi}_i)^2-\frac{1}{N} S_2\,, 
\end{equation}
and a symmetric traceless tensor (recall $\tilde{\delta}_{ij}:=\delta_{ij}-1/N$)
\begin{equation}
    Q_{ij}= \tilde{\phi}_i \tilde{\phi}_j- \frac{S_2}{N-1}\tilde{\delta}_{ij} + \frac{V_{2,j}-N \, V_{2,i} \tilde{\delta}_{ij}}{N-2}\,.
\end{equation}
Notice that $S_2$, $V_{2,i}$ and $Q_{ij}$ are the three irreps that appear in the tensor product of the fundamentals $\tilde{\phi}_i\times \tilde{\phi}_j$. The other two, $S_1$ and $V_{1,i}$, are formed by involving the singlet $\phi^T$. 

Consider a singlet bilinear operator $\mathcal{O}^T$ and a vector bilinear $\widetilde{\mathcal{O}}_{i}$ from the above five. Let them be normalized in a way 
\footnote{E.g. we may choose $\mathcal{O}^{T}=(2c_0^2(N-1)/N)^{-\frac 12} S_2$ and $\tilde{\mathcal{O}}_i=(2c_0^2(N-2)/N)^{-\frac 12}V_{2,i}$.} 
that the 2-point functions have the following forms
\begin{align}\label{mulcorr-new}
    &\braket{\mathcal{O}^{T}(\tau)\mathcal{O}^{T}(0)}=\frac{N }{\tau^{2\Delta_2}}\,,\\ \label{mulcorr2-new}&\braket{\widetilde{\mathcal{O}}_{i}(\tau)\widetilde{\mathcal{O}}_{j}(0)}=\frac{(\delta_{ij}-\frac{1}{N})}{\tau^{2\tilde\Delta_{2}}}\,.
\end{align}
Note that $\langle {\mathcal{O}}^T(\tau)\widetilde{\mathcal{O}}_{i}(0) \rangle$ vanishes by symmetry. 
In free theory $\Delta_2=\tilde\Delta_2=d-2$, but in presence of interactions $\Delta_2$ and $\tilde\Delta_2$  will acquire anomalous dimensions dependent on $N$. 

If $\Delta_2=\tilde\Delta_2+O(N)$ as $N\to 0$ then it natural to expect a logarithmic multiplet to form from ${\mathcal{O}}^T$ and $\widetilde{\mathcal{O}}$ in the same way as $\phi^T$ and $\tilde\phi$ before. In particular, the log partners should be $\mathcal{O}:=\widetilde{\mathcal{O}}_{1}+\frac 1N{\mathcal{O}}^T$ and $\mathcal{O}^T$. Following the steps that led to \eqref{log-phi} one will get the 2-point form
\be\label{biln-log}
\lim_{N\rightarrow0}\braket{\mathcal{O}(\tau)\mathcal{O}(0)}=\frac{2(\Delta_2'(N)-\tilde\Delta_2'(N)) \log \tau}{\tau^{2\Delta_{2}}}\Big|_{N=0}\,,
\ee
It is not obvious, however, which of the two singlets couples with which of the two vectors. For this we have to find the one loop corrected dimensions and identify $\Delta_2$ and $\tilde\Delta_2$.
This means we have to solve two mixing problems: 
one for the mixing of $S_1$ and $S_2$, and another for $V_{1,i}$ and $V_{2,i}$. 

The  one-loop renormalization of the bilinear defect singlets is the same computation as the defect beta functions \eqref{beta1} and \eqref{beta2}. 
The anomalous dimension matrix $\Gamma^{(S)}_{ij}=\partial \beta^{\text{1-loop}}_{h_i}/\partial_{h_j}$ (where $\beta^{\text{1-loop}}_{h_i}$ has the classical part $-\epsilon h_i$ removed) is given by 
\be
\Gamma^{(S)}=
\left(
\begin{array}{cc}
 \frac{\epsilon }{3}-\frac{h_1^*}{\pi } & -\frac{h_2^* (N-1)}{\pi } \\
 -\frac{h_2^*}{\pi } & -\frac{h_2^* (N-2)+h_1^*}{\pi } \\
\end{array}
\right)\,.
\ee

For the mixing problem of the two vectors it is simple to consider the operators $V'_{1,i}=\phi_i^2-\text{trace}$ and $V'_{2,i}=\phi_i\phi^T-\text{trace}$. It is easily seen that $V_{2,i}$ and $V_{1,i}$ from \eqref{vectors} are linear combinations of them. We may add $V'_{2,i}$ and $V'_{1,i}$  (for say $i=1$) to the defect as interactions, with respective couplings $t_1$ and $t_2$. Then computing their beta functions $\beta_{t_i}$ \footnote{For the beta functions of $t_1$ and $t_2$ one may look at the one-point functions of the two bulk operators $\big(\Phi_i^2-\text{trace}\big)$ and $\big(\Phi_i\sum_i\Phi-\text{trace}\big)$ and demand their finiteness. It is convenient to ignore all terms $O(t_it_j)$ since they do not contribute to $\Gamma^{(V)}$ (as we have $t_1=t_2=0$ at our fixed point). Then the computation is analogous to that of the singlets. }, one can obtain the anomalous dimension matrix $\Gamma^{(V)}_{ij}=\partial \beta^{\text{1-loop}}_{t_i}/\partial_{t_j}$ as
\be
\Gamma^{(V)}=
\left(
\begin{array}{cc}
 \frac{\epsilon }{3}-\frac{h_1^*}{\pi } & -\frac{h_2^* (N-2)}{2\pi } \\
 -\frac{h_2^*}{\pi } & -\frac{h_2^* (N-4)+2h_1^*}{2\pi } \\
\end{array}
\right)\,.
\ee

Let us now point out that $\Gamma^{(S)}=\Gamma^{(V)}+O(N)$ as $N\to 0$ as we had expected, and they have the eigenvalues $(7\pm \sqrt{7})/6$ (corresponding to $\Delta_2=2+\epsilon(1\pm \sqrt{7})/6$, see \eqref{DeltaAB}). 
For any $N$ the eigenvectors of $\Gamma^{(S)}$ give two distinct singlets with properties as in \eqref{mulcorr-new}, and those of $\Gamma^{(V)}$ give two distinct vectors with properties as in \eqref{mulcorr2-new}. 
We therefore obtain a pair of rank 2 logarithmic multiplets:  $(\mathcal{O},\mathcal{O}^T)_A$ and $(\mathcal{O},\mathcal{O}^T)_B$  with the dimensions $\Delta_A$ and $\Delta_B$ as in \eqref{DeltaAB}, respectively. The form of these operators are given by the eigenvectors of the mixing matrices. \\

Finally let us comment on how the log features discussed in this section appear in the disordered theory observables. The single replica field correlator in \eqref{mulcorr1} corresponds to a disorder averaged 2-point function $\overline{\langle\phi(\tau)\phi(0)\rangle}$ by the relation \eqref{replicaEq2}. The correlators of the bilinears are more subtle as they involve 2, 3 or even 4 distinct replicas. These objects will map to various disorder averaged 4-point functions (such as $\overline{\langle\phi(\tau_1)\rangle\langle\phi(\tau_2)\rangle\langle\phi(\tau_3)\rangle\langle\phi(0)\rangle\rangle}$,  $\overline{\langle\phi(\tau_1)\phi(0)\rangle\langle\phi(\tau_2)\rangle\langle\phi(\tau_3)\rangle\rangle}$, etc) with coincident pairs of external points. The logs identified in \eqref{biln-log} will thus appear at special limits in the linear combinations of these 4-point observables.    
\\ \\
{\bf Discussion:}
In this work we established the existence of interacting disordered fixed points of two dimensional surface defects in the bulk $4-\epsilon$ dimensional Wilson-Fisher CFT by working in the replica formalism. We found that the fixed points are not RG stable but can be reached by appropriately tuning the ordered coupling on the defect. We proceeded to compute one-loop anomalous dimensions of defect operators of the form $\phi_i$ and $\phi_i\phi_j$ and by organizing them into representations of $S_N$ and taking the replica limit $N\to 0$, we explicitly verified the emergence of logarithmic conformal multiplets in the defect spectrum.

A key unanswered question is whether there exist more general bulk scalar CFTs, like the $O(N)$ model or the hypercubic CFT, where the disordered fixed point is actually stable under RG. It would be interesting to generalize our analysis to a fully general scalar theory with quartic couplings, in the same spirit as \cite{Osborn:2017ucf,Rychkov:2018vya,Osborn:2020cnf,Bartlett-Tisdall:2025kcx}. If such fixed points exist, it is quite possible they can be experimentally realized particularly in the case of the $O(3)$ and/or cubic universality classes. Insights from a bootstrap-minded approach to logarithmic CFT could also play an important role in this analysis
\cite{Hogervorst:2016itc,Banerjee:2019uxo,Bhat:2025ygi,He:2025clu}.

Another important direction is to push the perturbative analysis further, to bare dimension 3 defect operators, where the universal displacement operator will appear. It would be interesting to understand properties such as energy transmission and reflection across non-unitary defects and whether well-established bounds on defect central charges, which involve displacement correlators, can be violated in this context. Additionally, it would be illuminating to study more complicated observables such as bulk two-point functions of composite operators which admit both a unitary bulk OPE decomposition and a non-unitary DOE expansion, in order to further probe the consistency of non-unitary submanifold dynamics with an interacting unitary bulk. 

Finally, let us emphasize that the nature of disorder in our setup makes it closely related to the RFIM. Although the RFIM is known to possess a Parisi-Sourlas SUSY fixed point, the critical exponents found numerically in $d=3$ and $4$ do not respect SUSY \cite{Fytas_2013, Fytas_2016}. Therefore, the model also has a non-SUSY fixed point, and a challenging problem is to understand its origin in RG and how (or if) it flows to the SUSY one. Since defects can realize both PS SUSY \cite{Ghosh:2025lzb} and non-SUSY fixed points (this work) they seem to be the perfect playground to understand the SUSY and non-SUSY interplay. 
\\ \\
{\bf Acknowledgements:}
We thank Diptarka Das, Parijat Dey, Bobby Ezhuthachan, Miguel Paulos, Giuseppe Policastro, Somnath Porey and Philline van Vliet for useful discussions. AA further thanks Andreia Gon\c{c}alves for continued inspiration. AA is funded by the European Union (ERC, FUNBOOTS, project number 101043588, PI Miguel Paulos). Views and opinions expressed are however those of the authors only and do not necessarily reflect those of the European Union or the European Research Council Executive Agency. Neither the European Union nor the granting authority can be held responsible for them. AK is supported by ANRF grant ANRF/ARG/2025/001338/PS. 

\vspace{10pt}

\begin{acknowledgments}

\end{acknowledgments}

\bibliographystyle{apsrev4-1}
\bibliography{letter}

\end{document}